\documentclass{appolb}

\usepackage{amsmath}
\usepackage{hyperref}
\usepackage{graphicx}
\usepackage{cite}
\usepackage{color}

\newcommand{\gmt}[1]{$g^2 \mu t = #1$}

\graphicspath{{plots/}}

\begin{document}
\eqsec
\title{A real-time lattice simulation of the thermalization of a gluon plasma: first results%
\thanks{Presented by B. Wagenbach at Excited QCD 2016: The 8th workshop on QCD at low energies and QCD at high temperatures \& finite densities, Costa da Caparica, Lisbon, Portugal, March 6\,-12, 2016.}
}
\author{Maximilian Attems$^a$, Owe Philipsen$^b$, Christian Sch\"afer$^b$,\\ Bj\"orn Wagenbach$^b$, Savvas Zafeiropoulos$^b$
\address{$^a$Departament de F\'\i sica Qu\`antica i Astrof\'\i sica \& Institut de Ci\`encies del Cosmos (ICC), Universitat de Barcelona, Mart\'{\i} i Franqu\`es 1, 08028 Barcelona, Spain}
\address{$^b$Institut f\"ur Theoretische Physik - Johann Wolfgang Goethe-Universit\"at, Max-von-Laue-Str. 1, 60438 Frankfurt am Main, Germany}
}

\maketitle


\begin{abstract}
To achieve an understanding of the thermalization of a quark-gluon plasma, starting from QCD without using model assumptions, is a formidable task. We study the early stage dynamics of a relativistic heavy ion collision in the framework of real-time simulations of classical Yang-Mills theory in a static box with the color glass condensate as initial condition. Our study generalizes a previous one by Fukushima and Gelis from $SU(2)$ to the realistic case of $SU(3)$. We calculate the chromo-electric and chromo-magnetic energy densities as well as the ratio of longitudinal and transverse pressure as a function of time as probes for thermalization. Our preliminary results on coarse lattices show the occurrence of Weibel instabilities prior to thermalization. 
\end{abstract}


\section{Introduction}

Experimental results from heavy-ion collisions at the Large Hadron Collider (LHC) and the Relativistic Heavy-Ion Collider (RHIC) give strong indications for the occurrence of a quark-gluon plasma (QGP). The successful description of the QGP, even at early times, by relativistic viscous hydrodynamics suggests a rather fast thermalization within a few fm \cite{Heinz:2001xi,Romatschke:2007mq}. Starting from the highly anisotropic scenario of two colliding ultrarelativistic nuclei, it is still an open question which processes are responsible for this fast thermalization. One candidate are color instabilities \cite{Mrowczynski:2016etf,Attems:2012js} such as the chromo-Weibel instability \cite{Arnold:2004ti,Bodeker:2005nv,Mueller:2005un,Mrowczynski:1996vh}, which is the analogue of the Weibel instability in electromagnetic plasmas \cite{Weibel:1959zz}.

In the limit of weak coupling, which is assumed for the present energy scales at LHC and RHIC, the two nuclei can be described in the Color Glass Condensate (CGC) framework \cite{Iancu:2003xm,Gelis:2010nm,Gelis:2012ri}. It provides an expansion of inclusive quantities, such as the expectation value of the energy-momentum tensor $T^{\mu\nu}$, in powers of the strong coupling $\alpha_s$, where the leading order is obtained by solving the Yang-Mills equations. We want to approach this problem from first principles by applying real-time lattice QCD techniques, approximating QCD by Yang-Mills theory in the classical limit. The classical approximation is justified by the high occupation of gluons in the non-equilibrium medium produced after the heavy-ion collision. In contrast to other works using Bjorken (expanding) coordinates \cite{Fukushima:2011nq}, our exploratory studies focus on finite, static box simulations.


\section{Color Glass Condensate}

The classical Yang-Mills equations for the gauge fields and their conjugate momenta can be solved using the framework of lattice gauge theory \cite{Ambjorn:1990wn}, where the continuum gauge fields $A_\mu(x)$ are represented in terms of lattice gauge fields $U_\mu(x) = \exp \big(igaA_\mu(x)\big)$. We follow the McLerran-Venugopalan model \cite{McLerran:1993ka,McLerran:1993ni,McLerran:1994vd} to construct the initial conditions for the lattice gauge fields $U$ and the chromo-electric fields $E$ within the CGC framework.

\subsection{Lattice gauge fields}\label{sec:cmf}
We set the longitudinal component of the collective gauge field $U_z = 1$ and construct the transverse components by merging the color fields from the two colliding nuclei, denoted by $U^{(1)}$ and $U^{(2)}$, by solving \cite{Krasnitz:1998ns}
\begin{align}\label{eq:37KV}
\text{Tr} \left\{ T^a \left[ (U^{(1)} + U^{(2)}) (1 + U^\dagger) - \text{h.c.} \right] \right\} = 0
\end{align}

\noindent for given $U^{(1)}$ and $U^{(2)}$, with $T^a$ 
the generators of $SU(3)$.

To get the color fields $U^{(1)}$ and $U^{(2)}$, one solves the Poisson equation $\Delta_L \Lambda^a_m(x) = - \rho^a_m(x)$ for nuclear color sources with the following Gaussian distribution:
\begin{align}
\big\langle \rho^a_m(x) \rho^b_n(y) \big\rangle = 
g^4 \mu^2 a^2 \delta_{mn} \delta^{ab} \delta(x - y)
\end{align} 

\noindent The color fields are then constructed from the obtained ``color potentials'' $\Lambda_m = T^a \Lambda_m^a$ via $U^m_i(x) = \exp\big(\mathrm{i}\Lambda_m(x)\big) \exp\big(\mathrm{i}\Lambda_m(x+\hat{\imath})\big)$, with $m = 1,2$ and $i=x,y$.

\subsection{Chromo-electric fields}
The transverse chromo-electric fields are set to zero and the longitudinal part is constructed out of the gauge fields $U$ via

\begin{align}
\begin{split}
E_z^a(x) &= -\frac{i}{2g} \text{Tr}\Bigg(T^a \sum_{i=x,y} \bigg\{\Big[U_i(x) - 1\Big]\Big[U_i^{(2)\dagger}(x) - U_i^{(1)\dagger}(x)\Big] - \text{h.c.}\\
&\hspace{0.45cm} + \Big[U_i^\dagger(x-\hat{\imath}) - 1\Big]\Big[U_i^{(2)\dagger}(x-\hat{\imath}) - U_i^{(1)\dagger}(x-\hat{\imath})\Big] - \text{h.c.} \bigg\}\Bigg)\;.
\end{split}
\end{align}

\subsection{Fluctuations}
In order to reach isotropization, fluctuations need to be added to the initial fields \cite{Romatschke:2005pm,Romatschke:2006nk}. This is done by a small random variation of the idealized $\delta$-function-like distribution of the initial chromo-electric fields. To this end, we add a fluctuation term $\delta E \sim \Delta$. We will discuss the effect of the seed $\Delta$ in \autoref{sec:nr}.


\section{Observables}\label{sec:ob}
To study the thermalization time, we observe the ratio of the longitudinal to the transverse pressure $P_L/P_T$ \cite{Gelis:2013rba}. The pressure components are related to the spatial diagonal elements of the energy-momentum tensor $T^{ii}$ \cite{Romatschke:2006nk} and the chromo-electric and chromo-magnetic longitudinal and transverse energy densities as:
\begin{align}
P_L(t) &= - T^z_z(t) = \epsilon_{E_T}(t) + \epsilon_{B_T}(t) - \epsilon_{E_L}(t) - \epsilon_{B_L}(t)\\
P_T(t) &= -\frac{1}{2} \big[ T^x_x(t) + T^y_y(t) \big] = \epsilon_{E_L}(t) + \epsilon_{B_L}(t)
\end{align}

The lattice expressions for the components of the chromo-electric and chromo-magnetic energy densities $H^{E/B}_i$, with $i = x,y,z$, are

\begin{align}
H^B_i(t,x) &= \frac{2 N_c}{g^2} \sum_{\substack{j<k \\ j,k \neq i}} \left[ 1 - \frac{1}{N_c} \text{Re}\text{Tr}~U_{jk}(x) \right]\\
H^E_i(t,x) &= \frac{1}{g^2} \text{Re}\text{Tr} \big[ E_i(x)E_i(x) \big] \label{eq:HEi}
\end{align}

\noindent from which we get the energy densities $\epsilon_{B/E_{L/T}}$ by averaging over the transverse lattice plane $(x,y)$ or the longitudinal axis $(z)$, respectively. 


\section{Numercial results}\label{sec:nr}

Our results are based on calculations on a $40^3$-lattice averaged over 30 independent measurements. We set the coupling constant $g=2$ and fix $g^2 \mu L = 120$ in accordance with the literature and conforming to the expectations of RHIC physics \cite{Fukushima:2011nq}. 

\begin{figure}[htb]
\begin{minipage}{0.49\textwidth}
\centerline{%
\hspace{0.1cm}\includegraphics[width=1.07\linewidth]{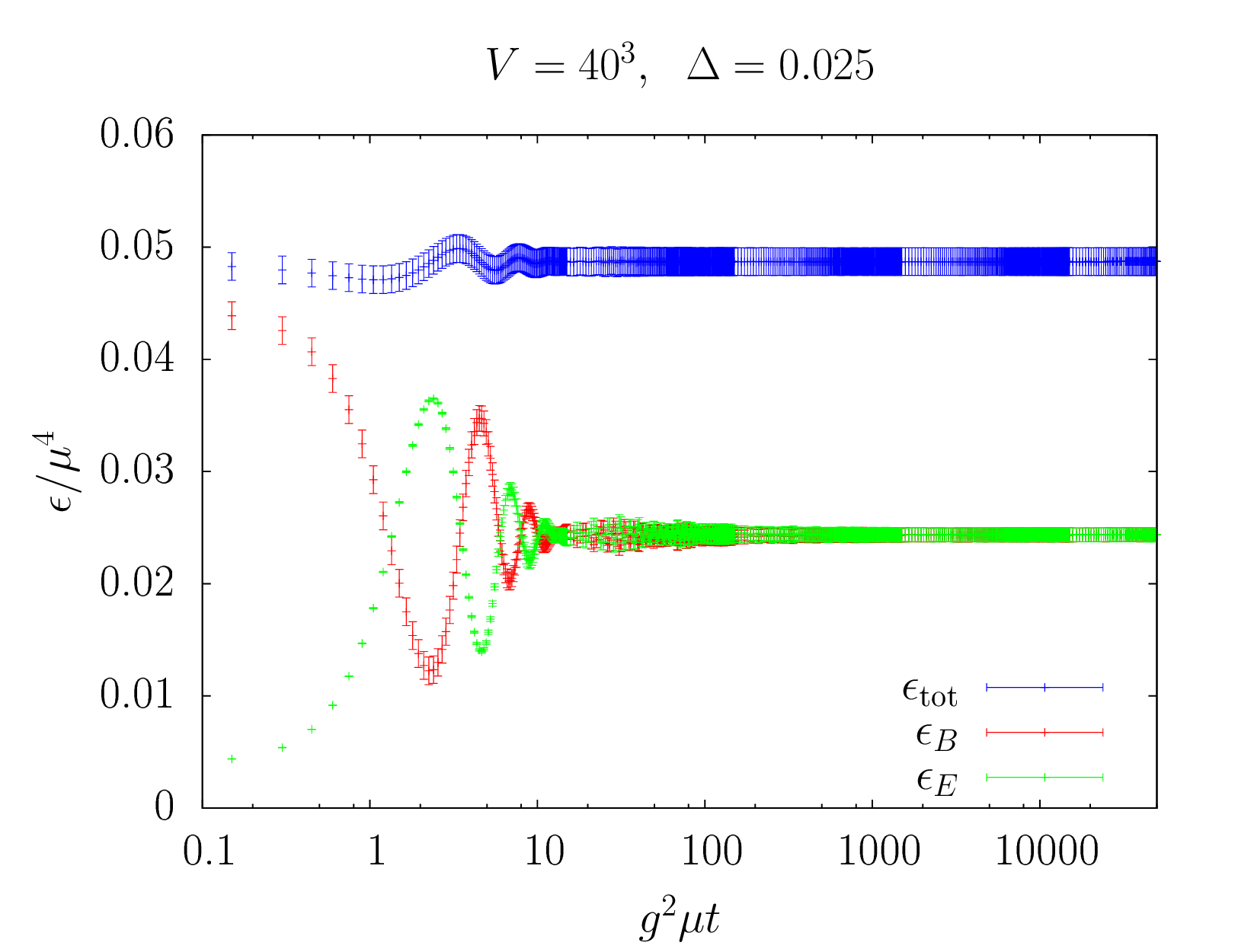}}
\end{minipage}
\begin{minipage}{0.49\textwidth}
\centerline{%
\hspace{0.5cm}\includegraphics[width=1.07\linewidth]{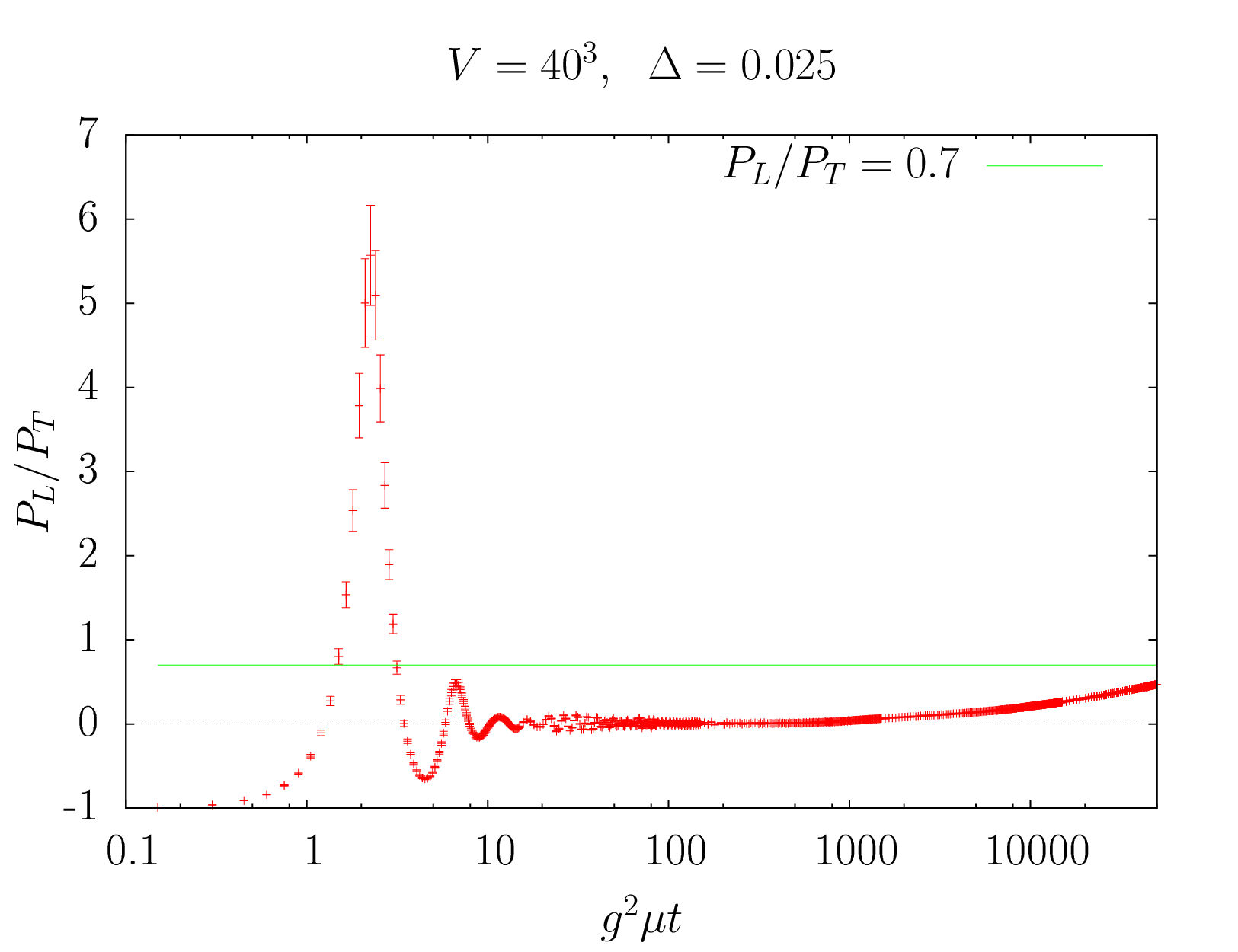}}
\end{minipage}
\caption{(Left) Magnetic energy density $\epsilon_B$, electric energy density $\epsilon_E$ and total energy density $\epsilon_\text{tot}$. (Right) Pressure ratio $P_L/P_T$ and hydrodynamization line at $P_L/P_T = 0.7$.}
\label{Fig:e_and_p}
\end{figure}

The value for the seed $\Delta$, which controls the fluctuations, is purely artificial. The higher it is, the faster the system thermalizes, but there is an upper bound: Since $\Delta$ only enters the electric part of the initial energy density, we have to monitor its influence on the initial total energy density, which should not significantly change, otherwise the classical approximation is no longer tenable. In our setup we chose $\Delta = 0.025$, which leads to a change of the initial value of the total energy density, which stays constant during the simulation, by less than $10\%$ (cf. \autoref{Fig:e_and_p} (left)).

Evidence for thermalization is provided by the pressure ratio $P_L/P_T$ approaching $1$. We are especially interested in the so-called hydrodynamization time, which is the time where hydrodynamic models should start to be applicable. \autoref{Fig:e_and_p} (right) shows the value of the pressure ratio, which slowly approaches the hydrodynamization level, which is commonly set to $P_L/P_T \approx 0.7$.

\begin{figure}[htb]
\begin{minipage}{0.32\textwidth}
\centerline{%
\hspace{.95cm}\includegraphics[width=1.4\linewidth]{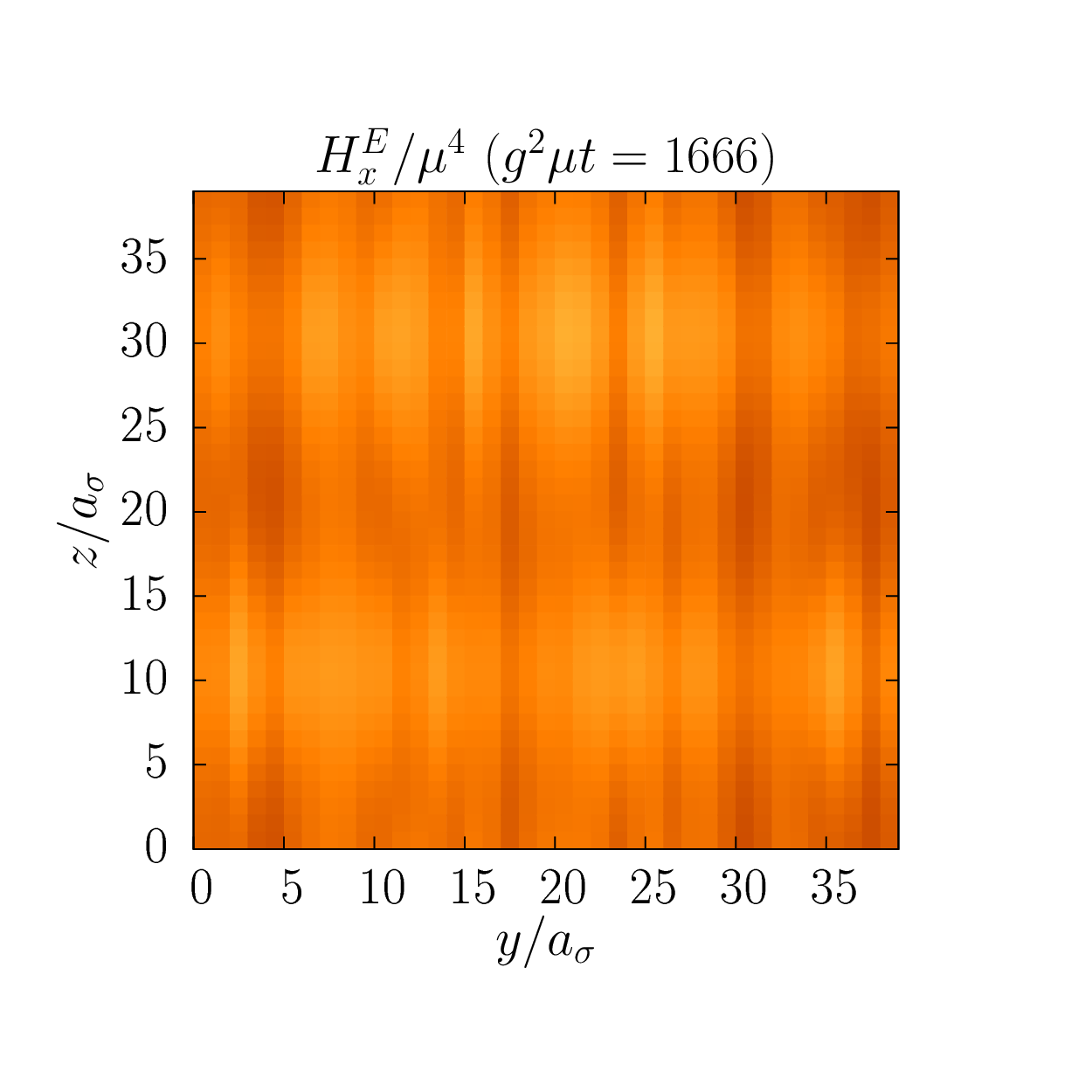}}
\end{minipage}
\begin{minipage}{0.32\textwidth}
\centerline{%
\hspace{0.05cm}\includegraphics[width=1.4\linewidth]{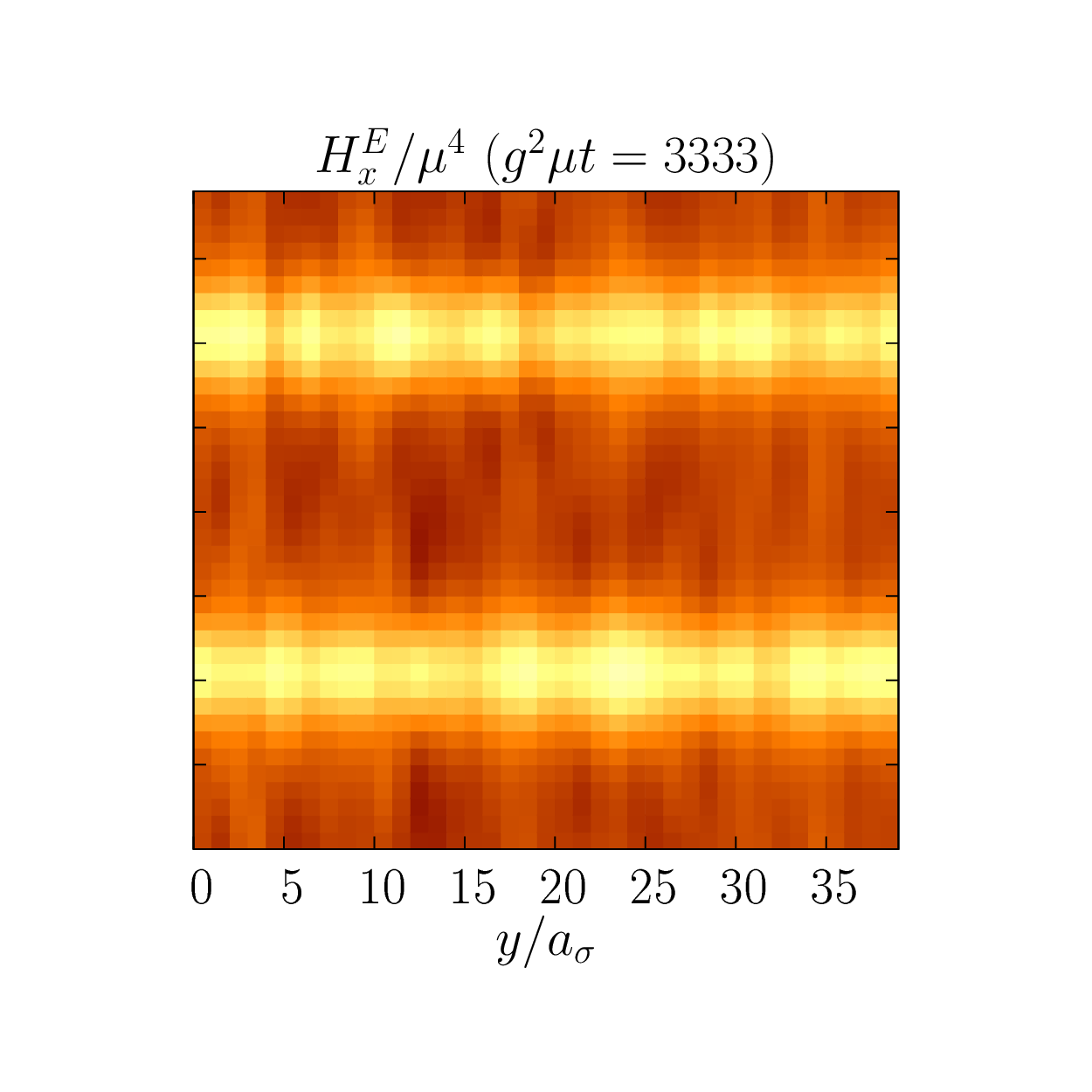}}
\end{minipage}
\begin{minipage}{0.34\textwidth}
\centerline{%
\includegraphics[width=1.317\linewidth]{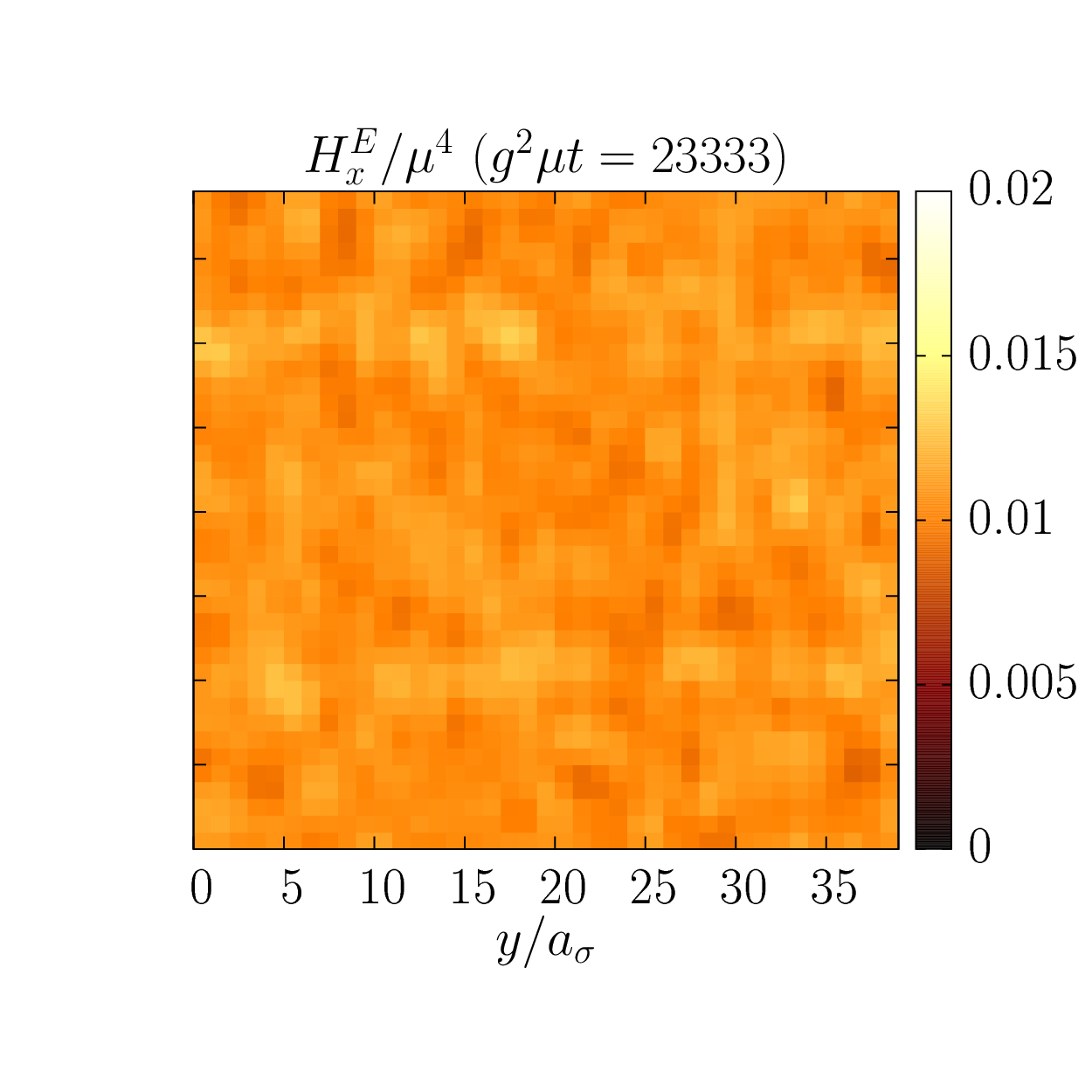}\hspace{1.1cm}}
\end{minipage}
\caption{The energy density $H^E_x$ averaged over the $x$-direction is plotted in the $(y,z)$-plane for different times. One can easily see the arising filamentation.}
\label{Fig:weibel}
\end{figure}

\autoref{Fig:weibel} shows the energy density $H^E_x$ (cf. \eqref{eq:HEi}) in the $(y,z)$-plane, where we averaged over the $x$-direction. Since we are only interested in the qualitative structure, we performed only one measurement here. This explains the vertical lines, which are a result of the initial fluctuations.

At \gmt{1666}, two filaments start emerging, which are clearly visible at \gmt{3333}. Later, at \gmt{23333} the filamentous structure is almost completely dissolved, which suggests an imminent thermalization. This characteristic filamentation is a strong indication for the presence of a chromo-Weibel instability.

\section{Conclusion}

We have provided evidence of the thermalization (hydrodynamization) of a color glass condensate such as the one created at relativistic heavy-ion collisions within classical $SU(3)$ Yang-Mills theory. To this end, we studied observables such as the energy density and its components and the longitudinal to transverse pressure ratio. By looking at components in a transverse-longitudinal plane, we were able to visualize the occurrence of chromo-Weibel instabilities represented by a clear filamentation of the energy density. The next steps will be an analysis of finite volume and discretization effects as well as studies of an expanding system.


\section{Acknowledgements}

The authors acknowledge fruitful discussions with B. Schenke and R. Venugopalan. M. A. is supported by the Marie Sk\l{}odowska-Curie Individual Fellowship 658574 FastTh. O. P and B. W. are supported by the Helmholtz International Center for FAIR within the LOEWE program of the State of Hesse. S. Z. is supported by the Humboldt Foundation. Calculations were done on the LOEWE-CSC at Geothe-University, the authors thank its administrative staff and the HPC-Hessen for programming support.


\end{document}